# Formation of quantum dots in MoS$_2$ with cryogenic Bi contacts and intrinsic Schottky barriers

Riku Tataka[1,2]*, Alka Sharma[3]*, Tomoya Johmen[1,2], Takeshi Kumasaka[1], Motoya Shinozaki[1,2], Yong P. Chen[3,4,5] and Tomohiro Otsuka[1,2,3,6,7,8]

[1] Research Institute of Electrical Communication, Tohoku University, 2-1-1 Katahira, Aoba-ku, Sendai 980-8577, Japan,

[2] Department of Electronic Engineering, Tohoku University, Aoba 6-6-05, Aramaki, Aoba-Ku, Sendai 980-8579, Japan

[3] WPI Advanced Institute for Materials Research, Tohoku University, 2-1-1 Katahira, Aoba-ku, Sendai 980-8577, Japan,

[4] Purdue Quantum Science and [5] Engineering Institute and Department of Physics and Astronomy, Purdue University, West Lafayette, Indiana 47907, USA

[6] Center for Spintronics Research Network, Tohoku University, 2-1-1 Katahira, Aoba-ku, Sendai 980-8577, Japan

[7] Center for Science and Innovation in Spintronics, Tohoku University, 2-1-1 Katahira, Aoba-ku, Sendai 980-8577, Japan

[8] Center for Emergent Matter Science, RIKEN, 2-1 Hirosawa, Wako, Saitama 351-0198, Japan

*Equal contribution

## Abstract

The recent advancement in two-dimensional (2D) materials-based quantum confinement has provided an opportunity to investigate and manipulate electron transport for quantum applications. However, the issues of metal/semiconductor interface effects create a hurdle to realize the artificial fabrication of the quantum dot and study the quantum transport. Here, we utilize the strategy of employing the semimetal for Ohmic contacts with transition metal dichalcogenides especially, multilayer MoS$_2$ and successfully fabricate the MoS$_2$-Bi based FET devices. We observe the Ohmic behavior in the MoS$_2$-Bi devices at cryogenic temperatures 4.2, 2.3 and 0.4 K. We also utilize intrinsic Schottky barriers formed at the interface between MoS$_2$ and Au for the gate electrodes to form and control quantum dots. We observed Coulomb diamonds in MoS$_2$ devices at cryogenic temperature. Our results of quantum transport in MoS$_2$ could serve as a stepping stone for investigating novel quantum effects such as spin-valley coupling and the manipulation of qubit systems.

The increasing demand for faster and more powerful computational techniques has led researchers to hit the atomic level and use quantum mechanical systems for information processing. Following quantum dot devices in graphene[1-8], the search for additional two-dimensional (2D) materials has led to the findings of transition metal dichalcogenides (TMDCs) for quantum devices, represented by molybdenum sulfide $MoS_2$[9-12]. These TMDC materials are bonded by van der Waals (vdW) force between individual layers[13]. In terms of electronics, carriers in TMDCs can be controlled by electric fields, which makes it promising for device applications such as field-effect transistors (FETs) [14-17]. The distinct properties of 2D TMDCs such as high mobility, spin and valley physics[18-21], long spin coherence time[22] and large spin-orbit coupling[23] make them potential candidates for quantum information and computation[24,25].

But the Schottky barriers between the contact metals and TMDCs induce a high contact resistance at cryogenic temperature[26]. The recent progress in the combination of $MoS_2$ and graphene in van der Waal heterostructures presented small barriers[27,28] and also other materials have been investigated[29,30]. Recently no barrier formation has been reported in monolayers of TMDCs with semimetal contacts at 77 K[31].

Here, we fabricate Bi/Au contacts which work at cryogenic temperature for quantum dot devices. The semimetal Bi/Au contacts showed a zero-barrier formation with multiple layers of $MoS_2$. We also realize gate electrodes utilizing intrinsic Schottky barriers formed at the interface between the pre-patterned Ti/Au fine gate electrodes and the transferred $MoS_2$. With these structures, we form quantum dots and observe Coulomb diamonds at cryogenic temperatures. These results will contribute to opening the newly unexplored research field focusing on electron transport in TMDCs quantum devices.

Multilayer MoS₂ thin flakes used in this experiment are exfoliated from the bulk crystal on PDMS and then transferred to Si/SiO₂ substrates. The thickness of the MoS₂ flakes is first identified by an optical microscope based on color contrast and further confirmed using the atomic force microscopy technique. We prepare MoS₂ devices in three-step processes. First, we prepared markers on Si/SiO₂ substrates using the photolithography technique. In the second step, we transferred MoS₂ flakes from PDMS to the Si/SiO₂ near the desired markers and followed by spin coating of the resist for the photolithography. Finally, in the third step, we prepare contacts with photolithography. The electron-beam (EB)-evaporation technique is used to deposit metal Ti - 10 nm and Au - 90 nm for the markers. The Bi – 40 nm and Au - 60 nm are deposited using thermal evaporation.

The controlled setup is employed to obtain electron transport measurements down to cryogenic temperatures such as 4.2, 2.3, and 0.4 K.

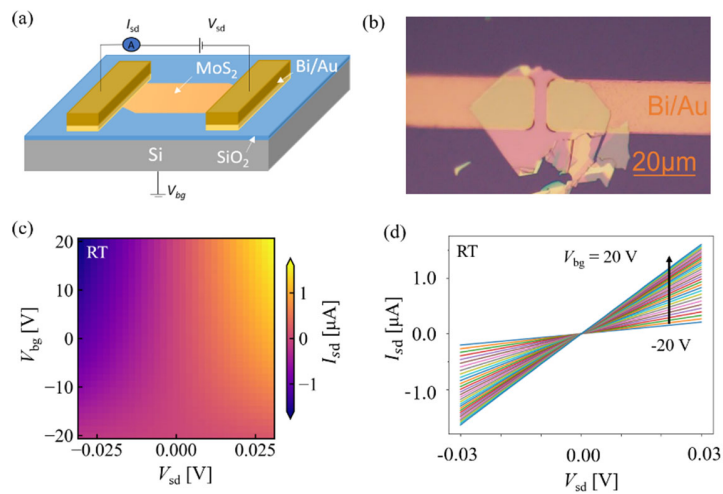

**Figure 1** (a) Schematic diagram of the MoS₂-Bi/Au based FET device structure. (b) The optical microscope image of the device. (c) Two-dimensional color map of the measured current as a function of source-drain bias and gate voltages in the MoS₂ device at RT. (d) *I-V* characteristics of the device at RT.

We start with the simple two probe geometry based MoS$_2$ FET device structure to investigate the contact resistance issues at the MoS$_2$ and metal interface, where Si under SiO$_2$ is used as a global back gate. The schematic of the MoS$_2$ based FET device geometry used for the two probe measurements is shown in Fig. 1(a). Usually, the contact resistance in TMDCs gets affected by the band structure of the semiconductor and metal work functions due to Fermi level pinning caused by metal- and disorder-induced gap states[32]. Here we have adapted the strategy of reducing the metal-induced gap states by employing Bi contacts and improving the contact resistance at cryogenic temperatures. By utilizing the Bi contacts, the gap-state saturation mechanism at the Bi–MoS$_2$ interface has been proposed and demonstrated in recent work[31]. We use semimetal Bi and form top source and drain contact electrodes. The thermal evaporator is utilized for the deposition of Bi at high vacuum conditions of 10$^{-6}$ Torr to minimize the interfacial contamination and then capped with Au to avoid oxidation. The microscope image of a fabricated device is shown in Fig. 1(b).

For the electrical characterization of the fabricated MoS$_2$ devices, we first performed the FET characterization at room temperature. A typical drain current $I_{sd}$ versus source bias voltage $V_{sd}$ for different back gate voltage $V_{bg}$ ranges from -20 V to 20 V characteristics is given in Fig. 1(c). With increasing $V_{bg}$, carriers are induced and finite conductance is observed. When the conduction channel is opened, the *I-V* characteristic shows linear relation in Fig. 1(d). Ohmic contacts are realized. The two-terminal resistance determined from the slope is 20 kΩ at $V_{bg}$ = 20 V and represents the n-channel conductance with Ohmic behavior similar to the earlier reports of MoS$_2$ transistors[33].

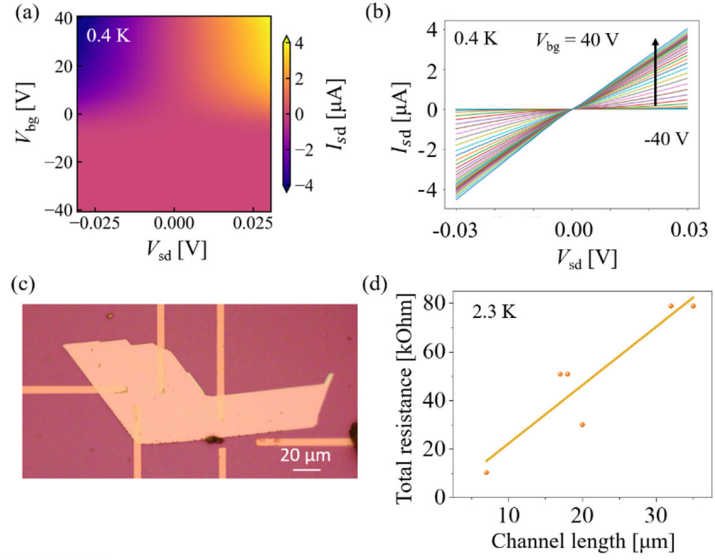

**Figure 2** (a) Two-dimensional color map of the measured current in the MoS$_2$ device at 0.4 K (b) *I-V* relation at different $V_g$ from -40 V to 40 V measured at 0.4 K. (c) The optical microscope image of the multi-contact device. (d) Channel length dependence of the toral resistance at 2.3K.

In the next step, we measure electron transport at low temperatures. The two-dimensional color map of the measured $I_{sd}$ as a function of $V_{sd}$ and $V_{bg}$ in the MoS$_2$ device at 0.4 K is given in Fig. 2(a). We can observe the opening of the conduction channel by applying positive $V_{bg}$. The pinch-off voltage of the FET is shifted to positive compared to the measurement at RT, probably due to suppression of thermally induced carriers. The *I-V* characteristic of the devices showed linear behavior in Fig. 2(b), which indicates good Ohmic contacts and zero Schottky behavior at the interface of the Bi and MoS$_2$ at the cryogenic temperature. Note that the probability that a device shows the Ohmic behavior is almost 100% at RT but the value decreases to roughly around 50% at cryogenic temperatures. This might be caused by imperfection or contamination of the interface in the fabrication process.

In addition, to investigate the channel length effect on the device resistance, we fabricated a multi-terminal MoS$_2$ FET device as shown in Fig. 2(c). We measure the resistance between several pairs of the contacts at 2.3K. $V_{bg}$ is fixed at 40 V. The measured resistance as a function of the channel length $L$ is shown in Fig. 2(d). Enhancement of the total resistance with the increase of $L$ is observed because of the increase in the channel resistance. The contact resistance is extracted as around 10k Ohm. Note that this value is a rough estimation, because the current path and the width is not precisely controlled in our device. With more controlled device, we will be able to obtain precise results. The Fermi level of a semimetal is close to the conduction band minimum of the semiconductor and hence suppresses the conduction band contributed metal-induced gap states, which leads to the pure contribution by the valence band[34]. Therefore, the easy charge transfer from Bi to $MoS_2$ is achieved.

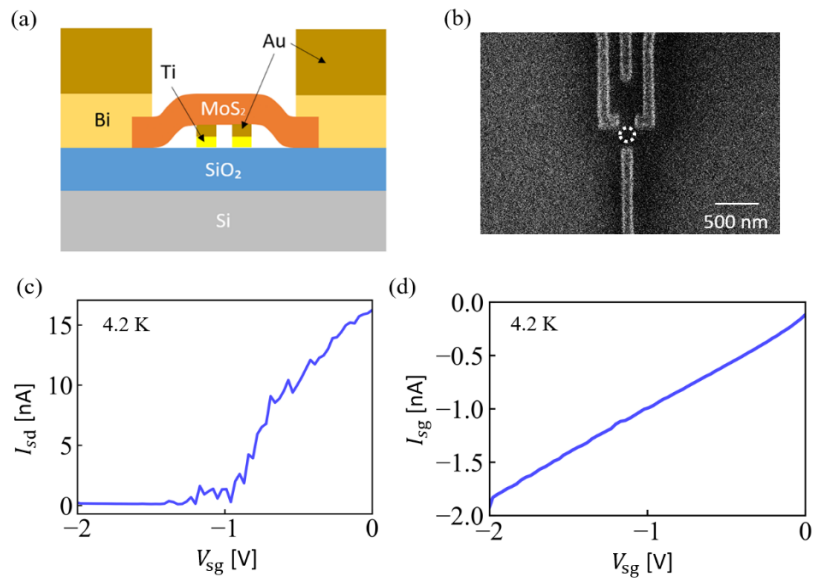

**Figure 3** (a) Schematic layer structure of the $MoS_2$ quantum dot device utilizing intrinsic Schottky barriers between Au and $MoS_2$. (b) SEM image of pre-patterned Ti/Au fine gates to form a quantum dot. (c) The observed pinch-off property by the surface gates. (d) The observed leak current through the surface gates including the leakage in the whole measurement setup. The current is suppressed in this voltage range.

Next, we develop gate electrodes to form confinement potential for quantum dots. We utilize intrinsic Schottky barriers formed at the interface between metal and $MoS_2$. Figure 3(a) shows the layer structure of the device. The gate electrodes are prepared using the EB lithography process on $Si/SiO_2$ substrate. The metals Ti - 10 nm and Au - 100 nm are deposited using the EB evaporation process. Figure 3(b) shows the scanning electron micrograph of the fine gates. Then, a multilayer

MoS$_2$ flake is placed on top of the fine gates by dry transfer technique[35] and treated with chloroform and hydrogen for the further cleaning of elvacite residues. MoS$_2$ touches the surface of Au gates and a Schottky barrier is formed. We can utilize the fine gates as gate electrodes to form the confinement potential of the quantum dot. After transferring MoS$_2$, top Bi - 40 nm and Au - 60 nm contacts are deposited using a thermal evaporation process. While conventional quantum devices need an insulator such as h-BN between TMDCs and surface gates[10], our devices don't have to use it. Because the Schottky barrier acts as an effective insulator. Note that this process is easier compared to the process with gate insulators and post-EB lithography after transferring MoS$_2$.

Figure 3(c) shows the observed $I_{sd}$ as a function of the surface gate voltage $V_{sg}$ at 0.4 K. $V_{sg}$ is applied on all of the surface gates. We set $V_{sd}$ = 80 V to induce carrier in the MoS$_2$ device. We can observe the pinch-off property of the conduction channel around -1.2 V. The Schottky gates work and deplete the conduction channel in MoS$_2$. Figure 3(d) shows the gate leakage current including the leakage of the whole measurement setup. The leakage current is less than 2 nA and mainly in the measurement circuit not in the device.

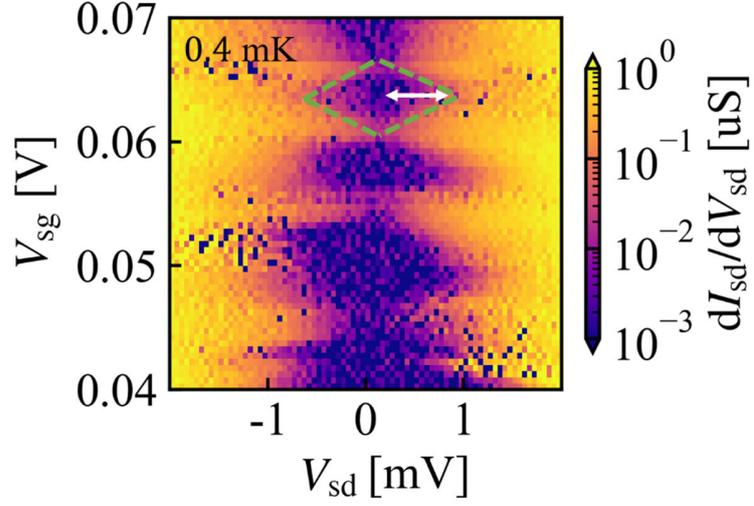

**Figure 4** Observed conductance as a function of the source-drain bias and the surface gate voltages. Coulomb diamonds are observed.

Next, we form a quantum dot using the surface fine gates. The two-dimensional map of the observed differential conductance $dI_{sd}/dV_{sd}$ as a function of $V_{sd}$ and $V_{sg}$ is given in Fig. 4. The conductance is suppressed around zero bias, and the width of the blocked region is modulated by $V_{sg}$. This corresponds to the Coulomb diamonds indicating the formation of a quantum dot. The size of the diamonds changes and the effect of the orbital level will be there. From the smallest size of the diamond, we extract the charging energy $E_c$ = 0.9 meV. This charging energy corresponds to a total capacitance of the quantum dot $C$ = 90 aF. We also extract the capacitance between surface gate and the quantum dot as 32 aF[36]. These values are consistent with the device geometry. The quantum dot will be formed at the gap of the surface gates (indicated by a circle in Fig. 3(b)). Surface charge impurities and bending of $MoS_2$ might also contribute to the confinement potential of the quantum dot.

## Conclusion

We have demonstrated $MoS_2$ quantum dot devices with cryogenic Bi contacts and Schottky gates utilizing intrinsic Schottky barrier between metal and $MoS_2$. We confirmed the operation of the contacts and gate electrodes. We observed Coulomb diamonds and the formation of a quantum dot at low temperatures and evaluated the parameters. Our results in layered $MoS_2$ could serve as a stepping stone for investigating novel quantum effects such as spin-valley coupling and the manipulation of qubit systems.

Note added. Recently, we became aware of a manuscript by R. T. K. Schock et al. about cryogenic Bi contact on MoS2 nanotubes and nanoribbons and the formation of quantum dots by global back gates[37].

## Acknowledgements

We thank S. Masubuchi, T. Machida, RIEC Fundamental Technology Center and the Laboratory for Nanoelectronics and Spintronics for fruitful discussions and technical supports. Part of this work is supported by MEXT Leading Initiative for Excellent Young Researchers, Grants-in-Aid for Scientific Research (21K18592), Fujikura Foundation Research Grant, Hattori Hokokai Foudation Research Grant, Kondo Zaidan Research Grant, Tanigawa Foundation Research Grant and FRiD Tohoku University.